\documentclass[a4paper, 10pt, twocolumn]{article}

\usepackage{graphicx}
\graphicspath{{./}{figures/}}

\usepackage[utf8]{inputenc}

\usepackage{microtype}
\makeatletter
\providecommand\MT@suspend@tagging{}
\providecommand\MT@resume@tagging{}
\makeatother

\usepackage[numbers,square]{natbib}

\usepackage{authblk}

\usepackage{booktabs}
\usepackage{color}
\usepackage{url}

\usepackage[acronym]{glossaries}
\usepackage{amssymb}
\usepackage{amsmath}
\usepackage{bm}
\usepackage{todonotes}
\usepackage{siunitx}
\usepackage[hidelinks]{hyperref}

\title{MAV-C: A Framework for the Joint Objective Estimation of Audio-Visual Complexity in Immersive Virtual Environments}
\date{}

\author[1,2]{Luca Resti\thanks{Corresponding author: luca.resti@york.ac.uk}}
\author[3]{Amelia Gully}
\author[2]{Michael McLoughlin}
\author[2]{Gavin Kearney}
\author[1]{Alena Denisova}

\affil[1]{Department of Computer Science, University of York, UK}
\affil[2]{AudioLab, School of Physics, Engineering and Technology, University of York, UK}
\affil[3]{Department of Language and Linguistic Science, University of York, UK}

\makeglossaries
\newacronym{QoE}{QoE}{Quality of Experience}
\newacronym{VR}{VR}{Virtual Reality}
\newacronym{AR}{AR}{Augmented Reality}
\newacronym{HCI}{HCI}{Human-Computer Interaction}
\newacronym{ML}{ML}{Machine Learning}
\newacronym{CAE}{CAE}{Complex Acoustic Environment}
\newacronym{VAE}{VAE}{Virtual Acoustic Environment}
\newacronym{HMD}{HMD}{Head-Mounted Display}
\newacronym{ASC}{ASC}{Acoustic Scene Complexity}
\newacronym{SASE}{SASE}{Spherical Acoustic Spatial Entropy}
\newacronym{MAV-C}{MAV-C}{Motion-Aware Audio-Visual Complexity Metric}
\newacronym{MSG}{MSG}{Multi-Scale Sobel Gradients}
\newacronym{MUC}{MUC}{Multi-Scale Unique Color}
\newacronym{AVC}{AVC}{Audio-Visual Complexity}
\newacronym{SE}{SE}{Spectral Entropy}
\newacronym{EE}{EE}{Energy Entropy}
\newacronym{EWM}{EWM}{Entropy Weight Method}
\newacronym{LTS}{LTS}{Log-Theoretical Scaling}
\newacronym{EWMA}{EWMA}{Exponentially Weighted Moving Average}
\newacronym{OF}{OF}{Optical Flow}
\newacronym{AWE}{AWE}{Amplitude-Weighted Entropy}
\newacronym{OFAT}{OFAT}{One-Factor-At-a-Time}
\newacronym{CoV}{CoV}{Coefficient of Variation}

\begin{document}

\maketitle

\begin{abstract}
We present the Motion-Aware Audio-Visual Complexity Metric (MAV-C), a reference-free framework for the joint objective estimation of audio-visual complexity. 
The metric combines entropy-based audio features (temporal, spectral, and spatial) with visual features 
(Sobel gradient magnitude, chromatic uniqueness, and optical flow) via a parametric fusion stage, producing a continuous joint complexity score $C_{AV}(t) \in \lbrack0,\!1\rbrack$. 
We validate MAV-C on two datasets: a controlled synthetic corpus (SYN) of stimuli with known signal characteristics and a naturalistic 
gameplay corpus (GAM) of 60 clips drawn from the SAFEPLAY-X dataset. On SYN, the metric exhibits strong validity: the audio score $C_A$ and visual score $C_V$ are each insensitive to 
changes in the opposite modality (CoV $<$ 0.003), the joint score $C_{AV}$ spans $\lbrack 0.00, 0.90 \rbrack$ across all parameter combinations, and single-axis feature 
sweeps produce monotone trajectories (Spearman $\rho$ up to 0.995). 
On GAM, $C_V$ differs significantly across content categories (Kruskal-Wallis $p = 0.021$) while $C_A$ does not, and the two sub-scores are uncorrelated ($r = 0.03$),
confirming they operate on independent signal dimensions. OFAT sensitivity analysis identifies a two-tier parameter hierarchy, 
with modality balance ($w_a$) and visual regularization ($\rho_v$) as most significant tunable parameters. Full subjective calibration is planned as future work.
\end{abstract}

\section{Introduction}

Many multimedia applications are inherently audiovisual: their perceptual impact is determined by the joint interaction of both modalities, not from either in isolation. 
Evaluating such content therefore requires assessment frameworks that
jointly process audio and visual signals rather than as independent streams.
Furthermore, many existing objective quality metrics often focus on signal-level distortions 
introduced by compression or transmission, but are blind to the informational and semantic 
demand of the content itself. 
In locally-rendered environments such as video games, encoding artefacts are absent or minimal, 
and existing metrics are therefore not well-suited to help practitioners assess the quality of audio-visual scenes.
Dense visual signals and multi-source acoustic environments can
saturate cognitive capacity and produce cross-modal interference
\cite{talsma2010multifaceted}, with consequences for user experience that distortion metrics
cannot capture.

The relationship between the structural complexity of audio and visual signals and human perception has been a growing area of research in recent years.
\gls{ASC} \cite{weisser2019complex} was found to be largely shaped by the spectro-temporal and spatial distribution of
sound events. Signal-based features such as energy and spectral entropy have recently been successfully used to
characterize \gls{ASC} objectively \cite{jorgensen2022acousticthesis}, and extended to account for the spatial distribution of 
sound sources through \gls{SASE} \cite{resti2026spherical}. In the visual domain, perceived complexity is
influenced by structural and chromatic signal variety \cite{kyle-davidson_characterising_2023}, and
\gls{MSG} and \gls{MUC} have been identified as interpretable features for predicting complexity
ratings \cite{saritas2025}. Both the subjective datasets and the objective models derived from
them \cite{weisser2019complex, kyle-davidson_characterising_2023, saritas2025} address a single
modality in isolation. Visual complexity research has focused on static images: to our knowledge,
no annotated dataset of perceived complexity exists for video (with audio), and no framework jointly
models audio and visual complexity over time on video signals. There is thus a significant gap for applications such as
video games, where \gls{ASC} and visual scene content are known to jointly influence perception
\cite{fichna2021effect}.

We propose \gls{MAV-C}, a reference-free framework that estimates \gls{AVC} by combining entropy-based audio features 
with visual complexity features mediated by a temporal motion component. The metric is parametric, with component weights designed
to be calibrated against human complexity ratings; no such joint audio-visual dataset currently
exists, so subjective validation is deferred to future work.

\section{The \gls{MAV-C} Architecture}
\label{sec:architecture}

\begin{figure*}[t!]
    \centering
    \includegraphics[width=2.1\columnwidth]{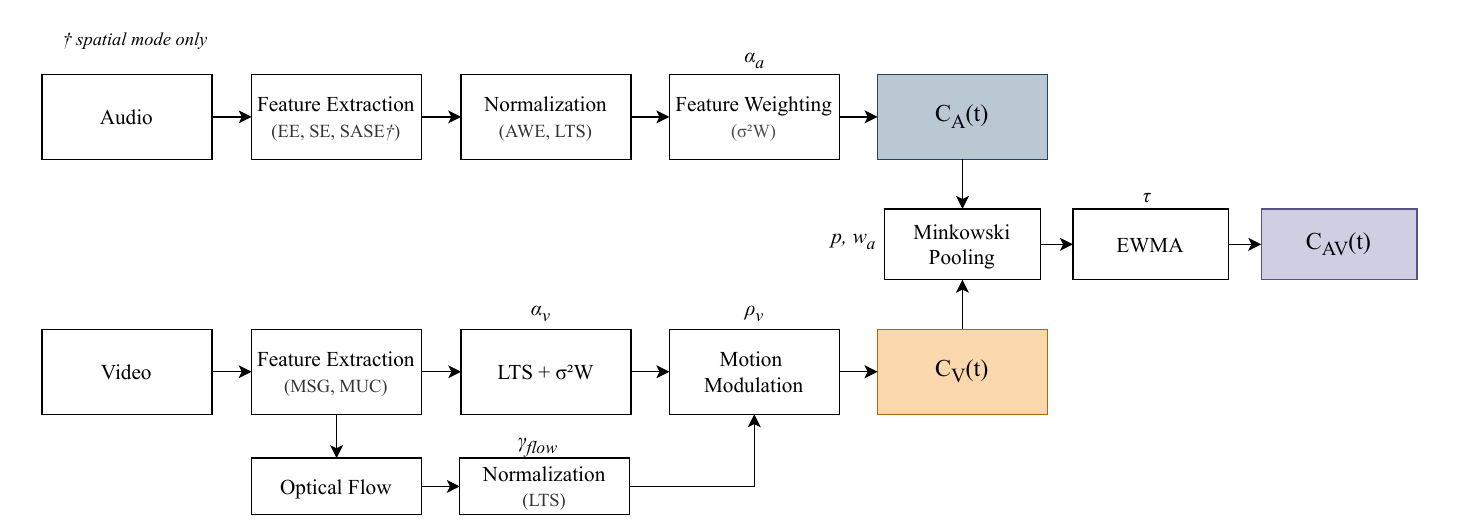}
    \caption{MAV-C block diagram. Parallel audio and visual pipelines produce sub-scores $C_A(t)$ and $C_V(t)$, fused into joint score $C_{AV}(t) \in [0,1]$. Annotated parameters are tunable.}
    \label{fig:mavc_block}
\end{figure*}

\gls{MAV-C} estimates joint audio-visual complexity as a continuous score $C_{AV}(t) \in [0, 1]$
from synchronized audio-visual stimuli. Each signal
feature is normalized against its physical upper bound (e.g. maximum possible entropy given the
number of bins), rather than against statistics of a particular dataset. This makes the scale
comparable across stimuli and resolutions without recalibration. The architecture has three
sequential stages: feature extraction, logarithmic range compression, and late fusion.

\subsection{Feature Extraction}
\label{subsec:feature_extraction}
Audio and visual features are extracted and aligned via a shared time grid.

\textbf{Audio Modality:} The audio sequence is analyzed at 100\,Hz with a 30\,ms
analysis window and 10\,ms hop length, yielding feature vector
$\mathbf{x}_A(t) \in \mathbb{R}^K$. All components are Shannon entropies across
different physical domains: time (\gls{EE}), frequency (\gls{SE}), and space (\gls{SASE}),
giving $\mathbf{x}_A(t) = [EE(t),\, SE(t),\, SASE(t)]^\top$. \gls{SASE} requires
audio source trajectory metadata and is optional: $K = 3$ in spatial mode and $K = 2$ in stereo mode when
such metadata is unavailable.

\gls{EE} quantifies the uniformity of energy distribution within an analysis frame
\cite{jorgensen2022acousticthesis}. Each 30\,ms frame is divided into $N_E = 10$ equal sub-frames.
Let $p_t(k)$ be the fractional signal power of sub-frame $k$, the temporal entropy is:
\begin{equation}
    EE(t) = -\sum_{k=1}^{N_E} p_t(k) \log_2 p_t(k).
\end{equation}
Impulsive transients concentrate energy in few sub-frames, yielding low \gls{EE}; sustained
broadband signals distribute energy uniformly, approaching the analytical bound of
$\log_2(10) \approx 3.32$.

\gls{SE} quantifies the uniformity of the acoustic power distribution across the frequency spectrum \cite{jorgensen2022acousticthesis}. 
From the Hamming-windowed STFT, letting $P_t(m)$ be the
normalized power at bin $m$ over the perceptually relevant band [16\,Hz,~10\,kHz]
($M = 300$ bins at 48\,kHz):
\begin{equation}
    SE(t) = -\sum_{m=1}^{M} P_t(m) \log_2 P_t(m)
\end{equation}
The theoretical maximum is $\log_2(300) \approx 8.23$.

Finally, \gls{SASE} is an optional feature that evaluates the spatial dispersion of acoustic
energy using source trajectory metadata \cite{resti2026spherical}. The spherical
acoustic map is partitioned into $N = 64$ equal-area discrete solid angles $\Omega_i$; letting
$I_t(\Omega_i)$ denote the normalized active intensity from sector $i$:
\begin{equation}
    SASE(t) = -\sum_{i=1}^{N} I_t(\Omega_i) \log_2 I_t(\Omega_i)
\end{equation}
Spatial complexity reaches its maximum of $\log_2(64) = 6.0$ when energy is equally distributed
across all sectors (perfectly diffuse field) and approaches its minimum when all energy arrives from a
single direction.

\textbf{Visual Modality:} Video is processed at its native frame rate, with features
merged onto the 100\,Hz audio grid. The visual feature space is composed of structural,
chromatic, and temporal components.

Let $\mathbf{I}(t) \in [0, 1]^{H \times W \times 3}$ denote a multi-channel video frame at time $t$.
Structural spatial complexity is quantified via \gls{MSG} \cite{saritas2025}. Applying horizontal
and vertical Sobel convolution operators $\mathbf{D}_x$ and $\mathbf{D}_y$ to the color channels
$c \in \{1, 2, 3\}$ at each of four spatial scales $s \in \{1, 2, 4, 8\}$, where $s$ is a
downsampling factor giving a frame of size $H/s \times W/s$, the per-pixel gradient magnitude
matrix $\mathbf{G}_{c,s} \in \mathbb{R}^{H_s \times W_s}$ is:
\begin{equation}
    \mathbf{G}_{c,s}(t) = \sqrt{(\mathbf{I}_{c,s}(t) * \mathbf{D}_x)^{\circ 2} + (\mathbf{I}_{c,s}(t) * \mathbf{D}_y)^{\circ 2}}
\end{equation}
where $*$ denotes 2D discrete convolution, $(\cdot)^{\circ 2}$ denotes element-wise squaring
(Hadamard power), and the $\sqrt{\cdot}$ is applied element-wise to yield a per-pixel
gradient magnitude map. The global \gls{MSG} score is the weighted summation of the spatial means
of these pixel-wise gradient magnitudes across all color channels and spatial scales:
\begin{equation}
\begin{split}
    MSG(t) &=\\ \sum_{s} w_s \Bigg( \frac{1}{3} \sum_{c=1}^{3} \frac{1}{H_s W_s} 
    & \sum_{x=1}^{W_s} \sum_{y=1}^{H_s} \mathbf{G}_{c,s}(x,y,t) \Bigg).
\end{split}
\end{equation}
The scale weights are $w_s \in \{0.4, 0.3, 0.2, 0.1\}$ for $s \in \{1, 2, 4, 8\}$, assigning
greater importance to full-resolution gradients \cite{saritas2025}. The analytical maximum,
derived from the Sobel kernel norm applied to a unit-range signal, is $\sqrt{20} \approx 4.47$.

Chromatic spatial complexity is evaluated via \gls{MUC} \cite{saritas2025}. At each scale $s$, the color
channels are uniformly quantized via bit-depth reduction, mapping the continuous color space to a
discrete set of states. Let $\mathcal{U}(\cdot)$ denote an operator that returns the cardinality
of the set of unique discrete vectors in a matrix. The \gls{MUC} score is the weighted summation
of unique quantized colors present at each scale, using the same weights as MSG
($w_s \in \{0.4, 0.3, 0.2, 0.1\}$):
\begin{equation}
    MUC(t) = \sum_{s} w_s |\mathcal{U}(\mathbf{I}_{q,s}(t))|.
\end{equation}
Because the number of unique colors is bounded by pixel count, the upper bound of \gls{MUC} depends
on native resolution and quantization depth. This maximum is computed per
stimulus at runtime and used for normalization, preserving the $[0,1]$ output range across
any input resolution.

The temporal visual complexity of a video sequence is quantified using dense optical flow. Let
$\mathbf{v}_{x,y}(t) \in \mathbb{R}^2$ represent the motion vector at pixel coordinates $(x,y)$.
The mean motion magnitude $OF(t)$ is the spatial mean of the $L_2$ norms of the flow vectors,
normalized by the frame diagonal $D = \sqrt{H^2 + W^2}$:
\begin{equation}
    OF(t) = \frac{1}{D} \left( \frac{1}{H \cdot W} \sum_{x=1}^{W} \sum_{y=1}^{H} \|\mathbf{v}_{x,y}(t)\|_2 \right).
\end{equation}
This restricts the relative optical flow sequence to the interval $[0, 1]$.

\subsection{Logarithmic Range Compression}
\label{subsec:range_compression}
To ensure the homogeneity of the parameter scales, \gls{MAV-C} projects the
raw feature vectors into a $[0, 1]$ subspace. Let $x_j(t)$ be the $j$-th raw feature extracted
at time $t$, and $x_{j,max}$ be its corresponding theoretical maximum.

This is implemented via \gls{LTS}, a nonlinear mapping $\mathcal{L}(\cdot)$ defined element-wise:
\begin{equation}
    \hat{x}_j(t) = \mathcal{L}(x_j(t)) = \frac{\log_2(1 + \gamma_j x_j(t))}{\log_2(1 + \gamma_j x_{j,max})}
    \label{eq:lts}
\end{equation}
where $\gamma_j > 0$ is a curvature parameter. The logarithmic form of $\mathcal{L}$ reflects the
Weber-Fechner law of sensory perception. For all auditory and visual features except optical flow, a fixed value $\gamma_j = 10$ is used.
This is not a tunable parameter in this study, as subjective
complexity ratings are required to meaningfully anchor the perceptual scale, which is deferred to future work.
Optical flow requires a separate parameter $\gamma_{flow}$. Normalization by the frame diagonal
$D$ makes $OF(t)$ resolution-independent, with the trade-off that even fast motion produces values
well below 0.1, since the mean per-pixel displacement is always a small fraction of $D$.
Applying $\gamma_j = 10$ to such a signal would utilize only the lowest portion of the $\mathcal{L}$ curve. 
$\gamma_{flow}$ compensates by increasing the curvature to spread the active flow range
across the full $[0,1]$ output. The sensitivity of this parameter is quantified in Section~\ref{subsec:sensitivity}.

\subsection{Modality Sub-Scores}
\label{subsec:sub_scores}
After log scaling, feature vectors are aggregated into independent auditory and
visual complexity terms, $C_A(t)$ and $C_V(t)$.

\textbf{Amplitude-Weighted Entropy:} Raw entropy features are not sensitive to signal level, whereby 
a quiet hum and a loud impulsive sound (e.g., an explosion) may yield near-identical \gls{EE}
and \gls{SE} values. To account for the perceptual salience of loud and complex sounds, both features
are scaled by an amplitude weight $w(t) \in [0,1]$, defined as the per-frame RMS mapped
linearly between the ITU-R BS.1770 silence gate (lower bound, $w=0$) and the EBU R128 true-peak
ceiling (upper bound, $w=1$) \cite{itu_r_bs1770, ebu_r128}. This yields the weighted vector
$\tilde{\mathbf{x}}_A(t)$, which guarantees $C_A(t) = 0$ for silent content.

\textbf{Acoustic complexity $C_A$:} The audio complexity score is the inner product of the
\gls{LTS}-processed and amplitude-weighted feature vector $\tilde{\mathbf{x}}_A(t)$ and a weight vector
 $\mathbf{w}_A \in \mathbb{R}^K$. 
$\mathbf{w}_A$ is computed via \emph{sigma-squared weighting} ($\sigma^2$W), which assigns
each feature a weight proportional to its within-clip variance, then regularizes toward a
uniform prior via parameter $\alpha_a \in [0,1]$:
\begin{equation}
    w_j = \frac{\alpha_a}{K} + (1-\alpha_a)\,\frac{\sigma_j^2}{\sum_{k=1}^{K} \sigma_k^2}, \quad \mathbf{1}^\top\mathbf{w}_A = 1
\end{equation}
where $\sigma_j^2$ is the within-clip variance of the $j$-th \gls{LTS}-compressed feature. At
$\alpha_a = 0$ weights are fully data-driven while at $\alpha_a = 1$ all features contribute equally.
The acoustic complexity score is then:
\begin{equation}
    C_A(t) = \mathbf{w}_A^\top \mathcal{L}(\tilde{\mathbf{x}}_A(t)).
\end{equation}
\gls{MAV-C} operates in two configurations: a stereo mode ($K=2$, using \gls{EE} and \gls{SE})
applicable to any stereo audio track, and a spatial mode ($K=3$, adding \gls{SASE}) when
directional source localization metadata is available.

For the visual domain, the static spatial complexity $S_V(t)$ is computed as the inner
product of the \gls{LTS}-processed \gls{MSG} and \gls{MUC} features using $\sigma^2$W
weights with visual regularization parameter $\alpha_v$:
\begin{equation}
    S_V(t) = \mathbf{w}_V^\top \left[\mathcal{L}(MSG(t)),\; \mathcal{L}(MUC(t))\right].^\top
\end{equation}
The final visual complexity score $C_V(t)$ scales $S_V(t)$ by a motion term, where
$\rho_v \in [0, 1]$ controls how much spatial complexity is retained when there is no motion:
\begin{equation}
    C_V(t) = S_V(t) \Big[ \rho_v + (1 - \rho_v) \mathcal{L}(OF(t)) \Big].
\end{equation}

\subsection{Feature Fusion and Cognitive Integration}
\label{subsec:fusion}
The modality sub-scores are fused into a joint complexity score $C_{AV_{raw}}(t)$.
Let $\mathbf{c}(t) = [C_A(t), C_V(t)]^\top \in [0,1]^2$ and let $\mathbf{w} = [w_a, w_v]^\top$
be the modality weight vector with $\|\mathbf{w}\|_1 = 1$. The fusion is a weighted
Minkowski pooling via the $L_p$ norm:
\begin{equation}
    C_{AV_{raw}}(t) = \left( w_a C_A(t)^p + w_v C_V(t)^p \right)^{\frac{1}{p}}
\end{equation}
where $p \ge 1$ is the pooling exponent. As
$p \to \infty$, the $L_p$ norm approaches the Chebyshev distance ($L_\infty$), biasing the joint
score in favor of the dominant modality. This behavior is motivated by attentional capture, as salient
events in one modality are known to dominate attention \cite{talsma2010multifaceted}.

Temporal integration is modeled as a causal first-order \gls{EWMA} filter applied to
$C_{AV_{raw}}$:
\begin{equation}
    C_{AV}(t) = \beta_t C_{AV}(t-1) + (1 - \beta_t) C_{AV_{raw}}(t).
\end{equation}
Coefficient $\beta_t = \exp(-\Delta t / \tau)$ depends on the sampling interval
$\Delta t$ and a time constant $\tau$ reflecting the integration time of human perception.

The sensitivity of all fusion parameters is characterized via \gls{OFAT} parameter sweeps in Section~\ref{subsec:sensitivity}.

\begin{figure*}[ht!]
    \centering
    \includegraphics[width=\textwidth]{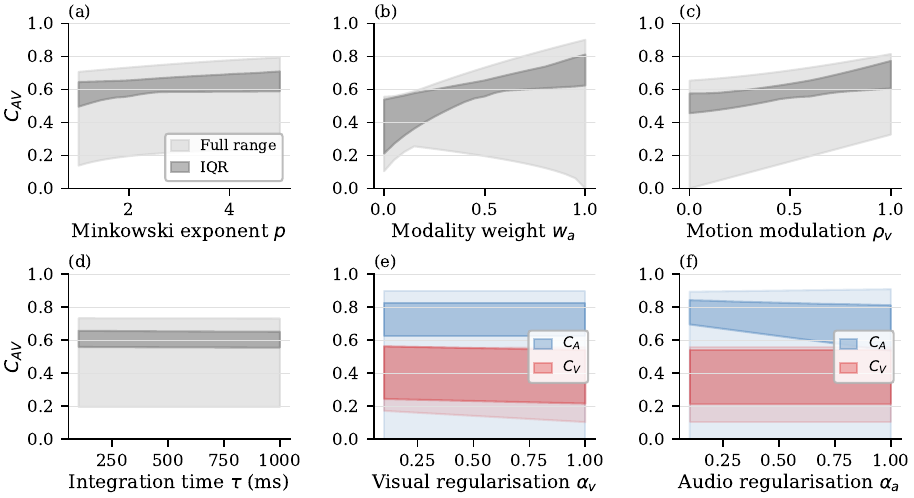}
    \caption{One-Factor-at-a-Time (OFAT) parameter sensitivity on SYN. Each panel varies one parameter while holding all others at their default values. 
    Panels~(a)-(d): $C_{AV}$; panels~(e)-(f): $C_A$ (blue) and $C_V$ (red). Light bands: full range across 18 conditions. 
    Dark bands: IQR. (a)-(c) primary fusion; (d) temporal integration; (e)-(f) adaptive weight regularization.}
    \label{fig:syn_param}
\end{figure*}

\section{Stimuli}
\label{sec:stimuli}

Two stimulus sets were used: a synthetic dataset (\textbf{SYN}) for
controlled verification of expected metric behavior, and a naturalistic video game gameplay dataset
(\textbf{GAM}) for real-world evaluation.

\textbf{SYN:} SYN comprises 18 audiovisual stimuli (15\,s, 30\,fps, 1280$\times$720;
48\,kHz audio), processed in spatial mode ($K=3$, including \gls{SASE}) with all frames
sampled. Audiovisual complexity was varied independently across different axes. Audio spans three
dimensions: spectral content (mix ratio of broadband noise to a pure sine tone), temporal
density (rectangular on/off gating duty cycle), and spatial distribution (number of virtual
sources $N \in [1,64]$ and their angular velocity $R \in [0^\circ,90^\circ\text{/s}]$);
spatial audio was rendered binaurally via SADIE~II KEMAR HRIRs \cite{Armstrong2018sadie}. Video
spans texture density (uniform noise block size from 1 to 64\,px) and motion speed
(0-15\,px/frame). The 18 conditions span floor and ceiling anchors, a midpoint baseline,
independence and \gls{SASE} boundary conditions, and single-feature continuous sweeps (see Table~\ref{tab:syn_conditions}).

\begin{table}[t!]
\centering
\caption{SYN conditions referenced in the results.}
\label{tab:syn_conditions}
\begin{tabular}{@{}lp{4.5cm}@{}}
\toprule
\textbf{Condition} & \textbf{Description} \\
\midrule
Floor        & Silence with a static, low-texture canvas \\
\midrule
Ceiling      & All axes simultaneously at maximum complexity \\
\midrule
Joint        & Audio and visual complexity increase together across time \\
\midrule
Audio-only   & Audio complexity increases, visual content held at low fixed level \\
\midrule
Independence & One modality sweeps its full range while the other is held fixed \\
\midrule
SASE boundary & Single localized source (floor) vs.\ fully distributed 64-source field (ceiling); $C_V$ held fixed \\
\midrule
Audio        & Full sweep of spectral, temporal, and spatial audio axes (\gls{EE}, \gls{SE}, \gls{SASE}) \\
\midrule
Texture      & Texture density swept from coarse (64\,px) to fine (1\,px blocks) \\
\midrule
Motion       & Motion speed swept from static (0) to fast (15\,px/frame) \\
\bottomrule
\end{tabular}
\end{table}

\textbf{GAM:} GAM comprises 60 gameplay clips (5\,s each) drawn from SAFEPLAY-X
\cite{singh2026safeplay}, a dataset originally assembled for violence
detection. Clips were stratified across three content categories (20 each): \textit{threat}
(suspense/stealth), \textit{assault} (active combat), and \textit{lethal} (culminating action).
These labels reflect narrative content type and carry no implied perceptual complexity ordering.
All clips were processed in stereo mode ($K=2$, \gls{EE} and \gls{SE} only) due to the lack of trajectory metadata.

\section{Results}
\label{sec:results}

\subsection{Default Parameters and Sensitivity}
\label{subsec:sensitivity}

\gls{MAV-C} has one \emph{fixed design constant}, the \gls{LTS} curvature $\gamma_j = 10$ (shared
across all features), and seven \emph{tunable fusion parameters}. Pending subjective calibration,
these are initialized to midpoints: $p = 2$, $w_a = 0.5$, $\alpha_a = 0.5$,
$\alpha_v = 1.0$, $\rho_v = 0.60$, $\gamma_{flow} = 500$, $\tau = 400$\,ms; this combination is
referred to as \emph{``default parameters''} throughout. To characterize how each parameter
independently affects the output, \gls{OFAT} sweeps are run across all GAM and SYN items, varying one
parameter at a time while holding the rest at default values.

\subsection{Synthetic Dataset}
\label{subsec:results_syn}

Figure~\ref{fig:syn_param} shows the \gls{OFAT} sensitivity across all 18 SYN conditions.
Across all parameter combinations, $C_{AV}$ spans $[0.00, 0.90]$. The light band shows the full range and the dark IQR band shows that most conditions cluster in the mid-range,
with the anchors driving the extremes.
At default parameters the anchor conditions yield a dynamic range of 0.54 (floor: 0.19, ceiling: 0.73). Single-axis sweeps
produce strongly monotone metric trajectories: Spearman $\rho = 0.995$ ($p{<}0.001$) for texture,
$\rho = 0.899$ ($p{<}0.001$) for audio, and $\rho = 0.896$ ($p{<}0.001$) for motion.

Modality independence is confirmed by the independence conditions (\gls{CoV} $<$ 0.003): when
one modality sweeps its full range the other sub-score varies by less than 0.010, confirming
each sub-score is driven solely by its own modality. \gls{SASE} boundary conditions show that $C_A$ rises from 0.624 (single localized source) to 0.786 (fully
diffuse 64-source field), while $C_V$ remains constant ($\Delta = 0.000$).
Across all 110 parameter combinations, the joint condition reaches a higher $C_{AV}$ ceiling than either unimodal condition alone, confirming that audio and visual complexity contribute 
independently and additively to $C_{AV}$.


\subsection{Gameplay Dataset}
\label{subsec:results_gam}

\begin{figure}[t!]
    \centering
    \includegraphics[width=1\columnwidth]{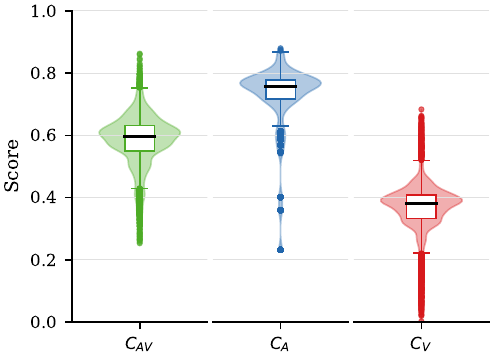}
    \caption{Distribution of per-clip mean $C_{AV}$, $C_A$, and $C_V$ across all 60 GAM clips and all parameter combinations (6600 observations: 60 clips $\times$ 110 parameter combinations). Violins show full density; boxes show median and IQR; outliers beyond 1.5$\times$IQR are plotted individually.}
    \label{fig:gam_distribution}
\end{figure}

Figure~\ref{fig:gam_distribution} shows the distribution of per-clip mean $C_{AV}$, $C_A$, and
$C_V$ across all 60 GAM clips and all 110 parameter combinations. Across all parameter combinations,
$C_{AV}$ spans $[0.00, 0.94]$. At default parameters the distribution narrows to $[0.35, 0.68]$
(median 0.60), situating naturalistic gameplay content in the upper-middle portion of the
metric's range and well within the $[0.19, 0.73]$ interval established by the SYN anchors,
confirming that the metric is neither saturated nor floored on naturalistic content.

\begin{figure}[t!]
    \centering
    \includegraphics[width=1\columnwidth]{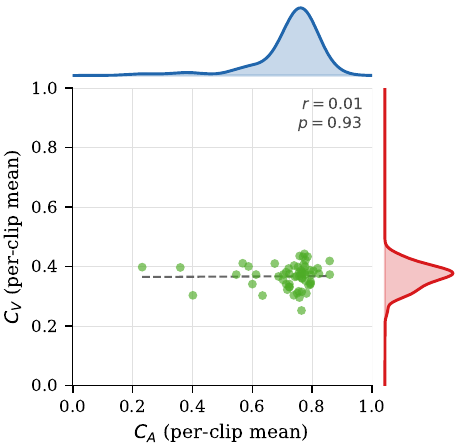}
    \caption{Per-clip mean $C_A$ vs.\ $C_V$ across 60 GAM clips, averaged over all parameter combinations. Dashed line: linear fit ($r = 0.03$, $p{=}0.824$). Marginal KDE curves for $C_A$ (blue, top) and $C_V$ (red, right). The near-zero correlation indicates that $C_A$ and $C_V$ vary independently on this corpus.}
    \label{fig:gam_scatter}
\end{figure}

Figure~\ref{fig:gam_scatter} shows the per-clip mean $C_A$ plotted against $C_V$. The two
sub-scores are statistically independent across the GAM corpus: Pearson $r = 0.03$
($p{=}0.824$; bootstrap 95\,\% CI $[-0.20,\, 0.33]$), Spearman $\rho = 0.13$ ($p{=}0.310$).
This indicates that $C_A$ and $C_V$ vary independently on this corpus.
A Kruskal-Wallis test across the three signal categories (threat, assault, lethal) reveals
significant separation in $C_V$ ($H = 7.70$, $p = 0.021$) but not in $C_A$ ($p = 0.437$),
consistent with the visual-dominant escalation characteristic of action gameplay: optical flow
and texture density increase with content intensity while the audio mix remains broadly
constant.

\section{Discussion}
\label{sec:discussion}

The SYN corpus confirms that the underlying architecture functions as intended: modality
sub-scores are mathematically isolated (\gls{CoV} $<$ 0.003) and respond monotonically to
targeted feature variations (Spearman $\rho \geq 0.88$). The GAM results support the argument 
for a joint metric: $C_V$ tracks narrative escalation across
content categories (KW $p = 0.021$) while $C_A$ does not ($p = 0.437$), and the
near-zero $C_A$/$C_V$ correlation ($r = 0.03$) confirms the two sub-scores capture
largely independent signal dimensions. A unimodal measure would therefore be blind to
whichever modality is omitted, regardless of that modality's contribution to scene complexity.

The appropriate modality balance varies with content type, task, and observer. The visual
dominance effect \cite{colavita1974, sinnett2007} suggests that visual salience often
captures attention preferentially, which may warrant adjusting $w_a$ for visually rich
content. \gls{MAV-C}'s parametric structure is designed to support perceptual calibration.
The sensitivity analysis (Figure~\ref{fig:syn_param}) reveals a two-tier
hierarchy across both datasets with $w_a$ and $\rho_v$ acting as primary levers (OFAT spans:
SYN 0.33 and 0.17; GAM 0.35 and 0.15). $p$, $\alpha_v$, and $\gamma_{flow}$ provide fine-tuning 
(GAM spans: 0.09, 0.03, and 0.01) and $\tau$ has negligible
effect on both ($<$0.01). The exception is $\alpha_a$, which spans 0.06 on SYN but
only 0.01 on GAM, indicating its influence is content-dependent and strongest for
stimuli with extreme spectral variation. With $C_{AV}$ spanning $[0.00, 0.94]$ across
all parameter combinations on GAM, the metric has the requisite flexibility to be tuned to specific 
subjective complexity ratings. The equal-weight default ($w_a = 0.5$)
is a conservative starting point pending such data.

\section{Limitations and Future Work}
\label{sec:limitations}

No subjective validation of \gls{MAV-C} was performed, as no joint audiovisual complexity
dataset (featuring video) with perceptual ratings currently exists. A subjective study to obtain 
such a dataset and tune metric parameters is planned. This will also
enable empirical determination of $\gamma_j$, currently fixed at a perceptually motivated value.
Corpus generalizability is also a constraint: GAM only features violent videogame scenes,
and the parameter hierarchy observed here may thus shift for other content types. On the signal
processing side, the spatial audio mode requires directional source metadata not always
available in production pipelines. Extending the framework to approximate source trajectories would broaden
its applicability. Dense optical flow is computationally intensive and may not capture
high-level motion salience, motivating the evaluation of lightweight alternatives.

\section*{Acknowledgments}
This work was supported by CoSTAR (Convergent Screen Technologies and Performance in Realtime)
Live Lab, funded by the Arts and Humanities Research Council, grant reference AH/Y001079/1 and
by the Engineering and Physical Sciences Research Council, grant reference EP/W524657/1.
This work was conducted in collaboration with Electronic Arts SEED.

\bibliographystyle{plainnat}

\bibliography{refs}

\end{document}